\documentclass[unnumsec,webpdf,contemporary,large]{oup-authoring-template}%

\usepackage{amsfonts,amsmath,amssymb,amsthm}
\usepackage{mathrsfs}
\usepackage{lmodern}
\usepackage{graphicx}
\usepackage{hyperref}

\setcitestyle{numbers}

\renewcommand{\Pr}{\operatorname{Pr}}
\DeclareMathOperator*{\argmax}{arg\,max}

\usepackage{draftwatermark}
\SetWatermarkText{Accepted}

\begin{document}

\journaltitle{PNAS Nexus}
\DOI{10.1093/pnasnexus/pgag085}
\copyrightyear{2026}
\pubyear{2026}
\access{Advance Access Publication Date: Day Month 2026}
\appnotes{Paper}

\firstpage{1}


\title[Tracing \emph{Sargassum} blooms to West Afica]{Tracing the origin of tropical North Atlantic \emph{Sargassum} blooms to West Africa}

\author[a,$\ast$]{Francisco Javier Beron-Vera}
\author[b]{Mar\'ia Josefina Olascoaga}
\author[c]{Philippe Miron}
\author[d]{Gage Bonner}

\address[a]{\orgdiv{Department of Atmospheric Sciences}, \orgname{Rosenstiel School of Marine, Atmospheric \& Earth Science}, \orgname{University of Miami}, \orgaddress{\street{4600 Rickenbacker Cswy.}, \postcode{33149}, \state{FL}, \country{USA}}}

\address[b]{\orgdiv{Department of Ocean Sciences}, \orgname{Rosenstiel School of Marine, Atmospheric \& Earth Science}, \orgname{University of Miami}, \orgaddress{\street{4600 Rickenbacker Cswy.}, \postcode{33149}, \state{FL}, \country{USA}}}

\address[c]{\orgdiv{Center for Ocean--Atmospheric Prediction Studies}, \orgname{Florida State University}, \orgaddress{\street{2000 Levy Ave.}, \postcode{32306}, \state{FL}, \country{USA}}}

\address[d]{\orgname{Morgridge Institute for Research}, \orgaddress{\street{330 N Orchard St}, \postcode{53715}, \state{WI}, \country{USA}}}

\corresp[$\ast$]{To whom correspondence should be addressed: \href{email:fberon@miami.edu}{fberon@miami.edu}}

\received{23}{4}{2025}
\accepted{2}{3}{2026}


\abstract{We simulate the dynamics of pelagic \emph{Sargassum} rafts as systems of finite-size floating particles, governed by a Maxey--Riley law with nonlinear elastic interactions. Using surface ocean currents and wind data from reanalysis systems for clump transport, we computed trajectories within a domain covering the tropical and subtropical north Atlantic. The subsequent motion is reduced using Ulam's discretization method into a time-inhomogeneous Markov chain that simulates a background \emph{Sargassum} concentration. Bayesian inversion, combined with nonautonomous transition path theory, was used to infer the origin of the first significant recorded bloom in the tropical North Atlantic, which unfolded in April 2011. Both methodologies independently identified the bloom's origin as near the West African coast, up to two years before it was detectable via satellite imagery on the basin's western side. This finding supports anecdotal evidence of \emph{Sargassum} strandings on the Ghanaian coast in 2009. Moreover, it correlates with unusual environmental conditions---such as increased nutrient loads from significant upwelling linked to a pronounced Dakar Ni\~na and Saharan dust deposition---that promote bloom proliferation. Additionally, it aligns with the observation that the species of \emph{Sargassum} in the 2011 bloom differ from those in the Sargasso Sea, which might otherwise be considered a natural origin.}


\boxedtext{The debated origin of \emph{Sargassum} seaweed blooms in the North Atlantic, first noted in 2011, is explored using a new model. We treat \emph{Sargassum} as elastic particle groups interacting with ocean currents and winds. By creating a time-inhomogeneous Markov chain, we estimate concentrations and trace the bloom origins to West Africa, consistent with 2009 reports from Ghana. Our results, using Bayesian inversion and transition path theory, diverge from the idea that blooms started in the Sargasso Sea, and they align more closely with environmental conditions linked to a Ni\~na-like event, where anomalous surface water cooling is accompanied by upwelling. This approach offers clarity on the differences in \emph{Sargassum} species origins.\\
\begin{quote}
    \emph{\textup{``}...en amaneciendo \textup{[}el viernes 21 de septiembre de 1492\textup{]} hallaron tanta yerba que parec\'ia ser la mar cuajada de ella, y ven\'ia del Oueste...\textup{'' (``}...at dawn \textup{[}on Friday, September 21, 1492\textup{]} they found so much grass that it looked like the sea was curdled with it, and it was coming from the west...\textup{'')}}
    
    Crist\'obal Col\'on (Christopher Colombus), \emph{Relaciones y cartas \textup{(}Accounts and letters\textup{)}}.
\end{quote}
}

\maketitle
\section{Introduction}

The translocation of \emph{Sargassum}, a significant species of brown algae within marine ecosystems, to the tropical North Atlantic region presents a compelling study of ocean--atmosphere interactions. Recent studies \cite{Johns-etal-20, Jouanno-etal-25} highlight the beginnings of the 2011 bloom \cite{Gower-etal-13}, which has recurred since and been termed the Great Atlantic \emph{Sargassum} Belt (GASB) \cite{Wang-etal-19}, situated in the Sargasso Sea, where abundance has historically been high \cite{Lapointe-86}. A significant anomaly in the North Atlantic Oscillation (NAO) \cite{Hurrell-etal-03} in 2009--2010 is believed to have disrupted the weak link between the subtropical and tropical North Atlantic \cite{Beron-etal-22-ADV}, allowing \emph{Sargassum} to be transported southward. However, evidence suggests a potential southern origin: anecdotal documentation of arrivals of \emph{Sargassum} in the coastal areas of Ghana in 2009 by Addicod and deGraft \cite{Addicod-deGraft-16}, overlooked by the oceanographic community, calls for consideration of an unexplored \emph{Sargassum} bloom origin. This paper aims to investigate this using a novel mechanistic model for the transport of \emph{Sargassum} and the application of a probabilistic framework which facilitates origin (source) inversion and the framing of efficient communication channels in the flow domain.

Specifically, in distinct departure from the ``leeway'' modeling approach \cite{Breivik-etal-13} of the previous studies \cite{Johns-etal-20, Jouanno-etal-25}, we simulate the dynamics of pelagic \emph{Sargassum} rafts through the modeling of such systems as finite-sized floating particles, or clumps, which evolve under a Maxey--Riley equation while subject to nonlinear elastic interactions.  The model, as developed in Bonner et al.\@~\cite{Bonner-etal-24}, is herein designated as eBOMB. The motion described by the Maxey--Riley equation within eBOMB is derived from the extension of the classic fluid mechanics equation \cite{Maxey-Riley-83} adapted to the oceanographic context \cite{Beron-etal-19-PoF, Olascoaga-etal-20}, with the ``inertial'' particles carried in flow created by the combined effects of ocean currents and winds. The Maxey--Riley oceanographic equation has been successfully validated both in situ \cite{Miron-etal-20-GRL} and under controlled laboratory conditions \cite{Miron-etal-20-PoF}. The elastic interactions within eBOMB offer enhancements over a previous model based on Hooke's law \cite{Beron-Miron-20}. For an overview of these and additional inquiries, the interested readers are directed to Beron-Vera \cite{Beron-21-ND}.

Using surface ocean currents and wind data from reanalysis systems, the flow carrying the \emph{Sargassum} clumps is determined. Trajectories are then computed by integrating eBOMB within a domain covering the tropical and subtropical regions of the North Atlantic's surface ocean. These trajectories' dynamics are subsequently reduced via Ulam's method \cite{Ulam-60} into a time-inhomogeneous Markov chain \cite{Norris-98, Bremaud-99}. Essentially, this represents a stochastic process where the probability of transitioning from state $i$ to state $j$ at time $t$ depends only on $i$, $j$, and $t$, and not on the history of the process before $t$. This simplification moves away from the generically chaotic nature of trajectories, creating a framework to determine the origin of \emph{Sargassum} blooms where probability densities play a central role \cite{Lasota-Mackey-94}. This form of reduction has been used within oceanographic contexts before \cite{Maximenko-etal-12, vanSebille-etal-12, Froyland-etal-14a} and more recently \cite{Miron-etal-17, Beron-etal-23-JPO}, especially in studying \emph{Sargassum} connectivity \cite{Beron-etal-22-ADV, Bonner-etal-23}. Bayesian inversion is then used to infer the origin of the first significant documented bloom in the tropical North Atlantic, observed in April 2011 adjacent to the western boundary of the basin. This is paired with a specialized application of the nonautonomous extension \cite{Helfmann-etal-20} of transition path theory (TPT) for stationary processes \cite{E-VandenEijnden-06}, which thoroughly characterizes nonequilibrium transitions while minimizing detours between distinct regions within a flow domain, thus identifying the most efficient communication pathways.  Applications of nonstationary TPT have so far been limited to toy-model Markov chains \cite{Helfmann-etal-20}.

The Bayesian and TPT inferences independently and consistently identified the origin of the 2011 bloom in near-coastal West Africa, almost two years before it was observed. This finding aligns with reports of \emph{Sargassum} washing ashore along the Ghanaian coast in 2009. Additionally, unusual environmental conditions linked to a strong Dakar Ni\~na \cite{Oettli-etal-16} were observed around 2009, creating favorable conditions for bloom development in the presence of baseline concentrations of \emph{Sargassum} species in the tropical North Atlantic, differing from that prevailing in the subtropics.

\section{Modeling \emph{Sargassum} Motion}

\paragraph{First Principles Framework.}

In the eBOMB model \cite{Bonner-etal-24}, a \emph{clump} of \emph{Sargassum} is the central element, influenced by ocean currents and winds. A clump essentially represents a group of \emph{Sargassum}, conceptualized as a solid sphere with a small radius. A network of such clumps that interact with each other is termed \emph{raft}, and it is assumed that the motion of a large amount of \emph{Sargassum} can be accurately depicted by the movement of these discrete clumps. At any given time, the entire \emph{Sargassum} collective being studied operates as a raft, although it might consist of numerous independent clump networks. The eBOMB model consists of three principal components: clump dynamics governed by coupled Maxey--Riley equations, nonlinear stiffness spring forces connecting clumps, and a biological model that governs clump growth and decay, which is not considered in this paper.

The choice to overlook biological interactions stems from their relatively limited understanding compared to physical interactions. This paper seeks to uncover the distant origins of observed blooms, which are expected to be mainly influenced by physical constraints. Although physiological changes in \emph{Sargassum} affect its biomass, they are expected to be less consequential than transport processes for this study, with ambient biogechemical conditions mainly influencing bloom initiation. The primary consideration is the life expectancy of \emph{Sargassum}, which is much better constrained than the intricate details of the life cycles.  The potential longevity of \emph{Sargassum} can reasonably be argued to extend over several years due to its clonal (vegetative) reproduction, with thallus sections persisting for multiple months before fragmentation, resulting in an effectively indefinite lifespan \cite{Godinez-etal-21}.

The physics of the eBOMB is influenced by several key parameters. Those describing inertial interactions are: $\alpha$, a dimensionless parameter representing windage; $R$, a dimensionless measure of the spherical clump's exposure to air; and $\tau$, the inertial response or Stokes' time. These parameters depend on the clump's density relative to water, $\delta \ge 1$, which is referred to as buoyancy. Additionally, $\tau$ depends on the clump's radius, $a$. Full expressions for these parameters can be found in the Supplementary Material SM1. Three additional parameters dictate the nonlinear elastic interactions among clumps: $\ell$, the natural length of the spring connecting neighboring clumps; $\kappa_0$, the stiffness amplitude of the spring; and $d$, the stiffness cutoff scale.

Let $\mathbf x = (x, y)$ represent, for ease of exposition, the position on a $\beta$-plane (cf.\@~Bonner et al.\@~\cite{Bonner-etal-24} for a formulation in full spherical geometry). The near-surface ocean velocity and wind at a position $\mathbf x$ and time $t$ are denoted by $\mathbf v(\mathbf x, t)$ and $\mathbf w(\mathbf x, t)$, respectively. Define
\begin{subequations}\label{eq:eBOMB}
\begin{equation}\label{eq:u}
    \mathbf u := (1 - \alpha) \mathbf v + \alpha \mathbf w.
\end{equation}
The trajectory $\mathbf x_m(t)$ of the $m$th clump of a raft obeys (cf.\ Supplementary Material SM2 for details):
\begin{equation} \label{eq:eBOMBxdot}
    \dot{\mathbf x}_m = \mathbf u\vert_m + \tau \mathbf u_\tau\vert_m  +  \tau \mathbf F_m, 
\end{equation}
where
\begin{equation}
    \mathbf u_\tau := R\frac{D\mathbf v}{Dt} + R \left( f + \tfrac{1}{3}\omega \right)\mathbf v^\perp -  \frac{D\mathbf u}{Dt} - \left(f + \tfrac{1}{3}R \omega \right) \mathbf u^\perp.
\end{equation}
Here,
\begin{equation}
    \frac{D\mathbf v}{Dt} = \partial_t\mathbf v + (\mathbf v\cdot\nabla)\mathbf v
\end{equation}
and similarly for $\frac{D\mathbf u}{Dt}$; $f = f_0 + \beta y$ is the Coriolis parameter (twice the local Earth's angular speed); $\perp$ is such that, for instance, $\mathbf x^\perp = (-y,x)$; $\omega = -\nabla\cdot\mathbf v^\perp$ is the vertical component of the ocean velocity's vorticity; and $\mathbf F_m$ denotes an external force.  For a raft comprising $M$ clumps, the resulting system comprises $2M$ first-order ordinary differential equations, coupled by $\mathbf{F}_m$. In contrast, the ``leeway'' modeling approach to \emph{Sargassum} transport \cite{Johns-etal-20, Jouanno-etal-25} simplifies the scenario by considering only the initial term on the right-hand side of equation \eqref{eq:eBOMBxdot}, leading to a set of decoupled equations.

The force $\mathbf F_m$ felt by the clump labeled $m$ is given by
\begin{equation} \label{eq:F-spring-def}
    \mathbf F_m = - \sum_{m' \in \operatorname{neighbor}(m,t)} \kappa(x_{mm'}) \left(1 - \frac{\ell}{x_{mm'}}\right) \mathbf x_{mm'},
\end{equation}
where $\mathbf x_{mm'} := \mathbf x_m - \mathbf x_{m'}$ and $x_{mm'} := |\mathbf x_{mm'}|$.  Here, $\operatorname{neighbor}(m,t)$ is equal to the set of indices of clumps that are connected to $i$ at time $t$ and
\begin{equation} \label{eq:k}
    \kappa(x_{mm'}) = \frac{\kappa_0}{\mathrm{e}^{(x_{mm'} - 2l)/d} + 1},
\end{equation}
\end{subequations}
where $d$ is taken small enough such that $\kappa(x_{mm'}) \approx \kappa_0$ for $0\le x_{mm'} \le 2\ell$ and $\kappa(x_{mm'}) \approx 0$ for $x_{mm'} > 2\ell$.  In this formulation, $\mathbf F_m$ acts as a restorative force maintaining the connection between clumps up to a specific distance, beyond which the clumps fully detach. 

\paragraph{Markov Chain Reduction.}

The deterministic framing of a \emph{Sargassum} bloom's origin using the eBOMB model \eqref{eq:eBOMB} is challenged by high sensitivity dependence of trajectories on initial conditions. These challenges may be mitigated by analyzing trajectory ensembles and creating histograms to illustrate the frequency of trajectory entries into designated regions within the two-dimensional ocean surface domain $\mathcal D$, where the motion takes place. Instead, we adopt a more appropriate method that \emph{reduces} the motion of \emph{Sargassum} clumps into an Markov chain via a discretization based on Ulam's method \cite{Ulam-60}. This approach shifts focus from individual trajectories to a probabilistic framework to determine the origin of a \emph{Sargassum} bloom, as developed following the discussion of the Markov chain reduction.

Consider $\mathcal D$ as a probability space wherein the Lebesgue measure is determined by area. Assume that there exists a \emph{nonautonomous}, or time-dependent, Perron--Frobenius operator $\mathscr P_{t_k,t_{k+1}}$ that \emph{transfers} the probability density of locating \emph{Sargassum} clumps within $\mathcal D$ at time $t_k = t_0 + k\Delta t$ to that at time $t_{k+1} = t_0+(k+1)\Delta t$, where $\Delta t>0$ and $k$ takes values in the \emph{discrete} set $\mathbb K := \{0,1,\dotsc,K-1\}$ so that $[t_0,t_{K-1}]$ represents a \emph{finite-time} interval. Furthermore, let $\mathscr P_{t_k,t_{k+1}}$ be characterized by a stochastic kernel $\mathscr K(\mathbf x, t_k;\mathbf y,t_{k+1}) \ge 0$ such that $\int_{\mathcal D} \mathscr K(\mathbf x, t_k;\mathbf y,t_{k+1})\,d\mathbf y = 1$ for all $\mathbf x\in\mathcal D$. Then, a probability density $f(\mathbf x)$, i.e., satisfying $\int_{\mathcal D} f(\mathbf x)\,d\mathbf x = 1$, at time $t_k$ evolves to time $t_{k+1}$ according to 
\begin{equation}
    \mathscr P_{t_k,t_{k+1}}f(\mathbf y) = \int_{\mathcal D} \mathscr K(\mathbf x, t_k;\mathbf y,t_{k+1}) f(\mathbf x)\,d\mathbf x.
\end{equation}
This density is the result of evolving $f(\mathbf x)$ under advection--diffusion dynamics with \emph{unsteady} drift, that is, a \emph{nonstationary} Markov (i.e., memoryless) random process. For a treatment of transfer operators and stochastic kernels, cf., e.g., Lasota and Mackey~\cite{Lasota-Mackey-94}. 

Ulam's method first involves partitioning $\mathcal D$ into a finite number of boxes $\{B_n\}_{1\le n\le N}$. Let $\mathcal D_N := \bigcup_{n=1}^NB_n$ denote the partition of $\mathcal D$.  The probability densities in $L^1(\mathcal D)$ are subsequently projected onto the finite-dimensional vector space $V_N$ spanned by $\smash{\big\{\text{area}(B_n)^{-1}\chi_{B_n}(\mathbf x)\big\}_{1\le n\le N}}$, where $\chi_\mathcal{A}(\mathbf x) = 1$ if $\mathbf x\in\mathcal A$ and zero otherwise is the indicator function of set $\mathcal A\subseteq\mathcal D$. Adapting the treatment of Miron et al.\@~\cite{Miron-etal-19-Chaos} for the autonomous case, one finds that the discrete action of $\mathscr P_{t_k,t_{k+1}} : L^1(\mathcal D) \circlearrowleft$ on $V_N$ is characterized by a matrix $\smash{\hat P(k) = \big(\hat P_{ij}(k)\big)_{i,j:B_i,B_j\in\mathcal D}} \in \mathbb R^{N\times N}$, where
\begin{align}
    \hat P_{ij}(k) &:=  \frac{\int_{B_i}\int_{B_j}\mathscr K(\mathbf x,t_k; \mathbf y,t_{k+1})\,d\mathbf x d\mathbf y}{\text{area}(B_i)} \nonumber\\ &= \Pr(\mathbf X_{k+1} \in B_j \mid \mathbf X_k \in B_i),
    \label{eq:P}
\end{align}
satisfying
\begin{equation}
    \hat P_{ij}(k) \ge 0\,\forall i,j:B_i,B_j\in\mathcal D_N,\, \sum_{j:B_j\in\mathcal D_N}\hat P_{ij}(k) = 1\,\forall i:B_i\in\mathcal D_N,
    \label{eq:stochastic}
\end{equation}
which specifies the proportion of probability mass in $B_i$ that flows to $B_j$ from $t_k$ to $t_{k+1}$, so
\begin{equation}
    \Pr(\mathbf X_{k+1} \in B_j) = \sum_{i:B_i\in\mathcal D_N} \hat P_{ij}(k)\Pr(\mathbf X_k \in B_i).
\end{equation}
Here, $\mathbf X_k$ is a time-$t_k$ $\mathcal D_N$-valued \emph{random variable} over an implicitly given probability space, where the probability measure is given by $\Pr$. In other words, $\mathbf X_k$ symbolizes random position in $\mathcal D_N$ at time $t_k$. Condition \eqref{eq:stochastic} implies that $\hat P(k)$ is a (row) stochastic matrix. Let $\mathbb D := \{1,2,\dots,N\}$ represent the set of indices of the domain's partition boxes.  By associating $\mathbf X_k\in B_i\in \mathcal D_N$ with $\hat X_k = i\in \mathbb D$, the matrix $\hat P(k)$ serves as the \emph{nonautonomous transition matrix} of the sequence $\{\hat X_k\}_{k\in\mathbb K}$ of random variables with values in the countable state space $\mathbb D$, which represents a \emph{one-sided discrete finite-time inhomogeneous Markov chain}.  For standard references on Markov chains, cf., e.g., Norris \cite{Norris-98} and Bremaud \cite{Bremaud-99}. A recent discussion of the inhomogeneous case with time taking discrete values in a finite interval is found in Helfmann et al.\@~\cite{Helfmann-etal-20}.

Let $\{\mathbf x_m(t)\}_{1\le m\le M}$ be a finite collection of \emph{Sargassum} clump trajectories, as produced by integrating \eqref{eq:eBOMB} over $t \in [t_0,t_{K-1}]$, such that they visit every box of $\mathcal D_N$. Each sampled trajectory set at time increment $\Delta t$, $\{\mathbf x_m(t_k)\}_{1\le m\le M}$, serves as an observation for $\mathbf X_k$. This enables the approximation of $P_{ij}(k)$ by counting the transitions between the partition boxes \cite{Miron-etal-19-JPO}:
\begin{equation}
    \hat P_{ij}(k) \approx \frac{\sum_{m=1}^M\chi_{B_i}\big(\mathbf x_m(t_k)\big)\chi_{B_j}\big(\mathbf x_m(t_{k+1})\big)}{\sum_{m=1}^M\chi_{B_i}\big(\mathbf x_m(t_k)\big)}.
\end{equation}
To ensure the stochasticity of $\hat P(k)$ when $\mathcal D$ is \emph{open}, as when it is taken to represent a domain of the surface ocean with no boundaries, we substitute $\hat P(k)$ with
\begin{subequations}\label{eq:Pext}
\begin{equation}
    P(k) := \begin{pmatrix} \hat P(k) & P^{\mathbb D\to\omega}(k)\\ P^{\mathbb D\leftarrow\omega}(k) & 0\end{pmatrix} \in \mathbb R^{(N+1)\times(N+1)}.
\end{equation}
Here, $\omega := \{N+1\}$ denotes a virtual state, called a \emph{two-way} nirvana state, which absorbs, at every time step, any probability imbalance in $\mathbb D$ and sends it back to the chain \cite{Miron-etal-21-Chaos}.  More specifically, 
\begin{equation}
    P^{\mathbb D\to\omega}(k) = \Bigg(1 - \sum_{j\in \mathbb D}\hat P_{ij}(k)\Bigg)_{i\in \mathbb D} \in \mathbb R^{N\times 1}
\end{equation}
gives the \emph{outflow} from $\mathbb D$ to $\omega$, while $P^{\mathbb D\leftarrow\omega}(k) \in \mathbb R^{1\times N}$ with entries that sum to one gives the \emph{inflow} from $\omega$ to $\mathbb D$.  The stochastic matrix $P(k)$ acts as the transition matrix for the Markov chain $\{X_k\}_{k\in\mathbb K}$, where $X_k$ takes values in the \emph{extended} state space $\mathbb D\bigcup \omega$.  There are several ways to model the inflow. In a previous application involving modeling the movement of \emph{Sargassum} using satellite-tracked buoys \cite{Beron-etal-22-ADV}, data on reentry from trajectories outside of $\mathcal D_N$ was utilized. In this paper, we express the inflow from $\omega$ to $\mathbb{D}$ as:
\begin{equation}
    P^{\mathbb D\leftarrow\omega}(k) = \left(\frac{1}{N}\right)_{i\in\mathbb D}, 
\end{equation}
\end{subequations}
simulating a background presence of \emph{Sargassum} on the ocean surface, as supported by the observations discussed below.  We note that satellite algorithms cannot detect low concentrations of \emph{Sargassum, such as individual clumps. They are only visible when blooms become sufficiently dense, which is important context for interpreting the main results of the paper.}

\section{Inferring the Origin of a \emph{Sargassum} Bloom}

\paragraph{Bayesian Inference.}

Let $B \subset \mathbb D$ represent the set of indices of the boxes that partition a region $\mathcal B \subset \mathcal D$ where a high concentration of \emph{Sargassum}, referred to as a bloom, has been observed at time $t_{k_B}$, where $k_B \in \mathbb K\setminus\!\{0\}$. Such a bloom has originated somewhere in the complement of $\mathcal B$, a remote location that we seek to frame.  Denote the set of indices of the partition of $\bar{\mathcal B} := \mathcal D\setminus \mathcal B$ by $\bar B$.  Motivated by the analysis of Miron et al.\@~\cite{Miron-etal-19-Chaos}, we call $T^B$ the \emph{random} time taken by the Markov chain to \emph{first hit} $B$, viz.,
\begin{equation}
    T^B := \inf_{k\in\mathbb K}\big\{t_k : X_k \in B\}.
    \label{eq:hit}
\end{equation}
Consider the computation of the probability that the chain is in $B$ at time $t_k$, conditioned on the chain's initial state, by pushing forward a probability vector supported on the initial state $\bar b$ while recording the probability value in $B$:
\begin{equation}
    p_{B,\bar b}(k) := \Pr(X_k \in B \mid X_0 = \bar b) = \sum_{b\in B}\left(\prod_{l = 0}^k P(l)\right)_{\bar bb}.
    \label{eq:p}
\end{equation}
Provided that $B$ is \emph{absorbing}, it follows that $p_{B,\bar b}(k) = \Pr(T^B \leq t_k\mid X_0 = \bar b)$. This is because, by definition, each trajectory that visits $B$ has a first visit at some time $t_k$ and subsequently remains in $B$. Since the first visit events $\{T^B = t_l\}_{l\in\mathbb K}$ are mutually exclusive, it follows that $p_{B, \bar{b}}(k) = \sum_{l = 0 }^{k} \Pr(T^B = t_l\mid X_0 = \bar b)$. Then, by replacing $k$ with $k - 1$ and subtracting each equation, we obtain:
\begin{align}
    p(t_k\mid\bar b) &:= \Pr(T^B = t_k\mid X_0 = \bar b) \nonumber\\ &= \begin{cases} p_{B,\bar b}(k) = 0 & \text{if } k = 0,\\ p_{B,\bar b}(k) - p_{B,\bar b}(k-1) & \text{if } k\in\mathbb K\setminus\!\{0\}.\end{cases}
    \label{eq:hit-prob}
\end{align} 
For each clump, we \emph{observe} two random quantities: the box partition $B$ of the ocean surface region $\mathcal B$ where \emph{Sargassum} clumps are highly concentrated, and the time $t_{k^B}$ when this concentration is recorded.  Let the corresponding random variables be called $X_*$ and $T_*$. Note that if $b$ is absorbing, then $T^B \le t_{K-1}$ and from \eqref{eq:hit} we have that $X_* \in B$. Consequently, the events $\{X_* \in B,\,T_* = t_{k^B}\}$ and $\{T^B = t_{k^B}\}$ are equivalent. Thus, given that the chain is in $\bar b$ at time $t_0$, the probability that the random variables jointly assume their observed values
\begin{align}
     \Pr(X_* \in B,\,T_* = t_{k^B}\mid X_0 = \bar b) &\equiv \Pr(T^B = t_{k^B}\mid X_0 = \bar b) \nonumber\\ &= p(t_{k^B}\mid\bar b),
\end{align}
which can be computed using \eqref{eq:hit-prob}. Bayesian inversion \cite{Bolstad-Curran-16} aims to determine a probabilistic depiction of $\bar b$ (``bloom's origin'') based on the observed bloom time ($t_{k^B}$) and its location (``$B$''). By Bayes' theorem, the \emph{posterior distribution} of $\bar b$, or the probability distribution of $\bar b$ after observing $t_{k^B}$, is computed as
\begin{equation}
  p(\bar b\mid t_{k^B}) \propto p(t_{k^B}\mid\bar b)\, p(\bar b),
  \label{eq:post}
\end{equation}
where $p(\bar b)$ denotes the \emph{prior distribution} of $\bar b$, which represents its existing knowledge before observing any data. Finally, \emph{maximum likelihood estimator} of the ``location of the bloom's origin'' is given by
\begin{equation}
    \hat b := \argmax_{\bar b}p(\bar b\mid t_{k^B}).    
\end{equation}

\paragraph{Transition Path Theory Inference.}

The extension of transition path theory (TPT) \cite{VandenEijnden-06, E-VandenEijnden-06, Metzner-etal-09} to nonautonomous finite-time dynamics, as developed in Helfmann et al.\@~\cite{Helfmann-etal-20}, enables a rigorous probabilistic study of \emph{nonstationary} transitions between two disjoint states, designated as source and target, in $S := \mathbb D\bigcup\omega$.  These are time-dependent transitions distinguished by their directness, characterized by the minimal occurrence of detours.  More specifically, a transition path is a trajectory that runs from the source state to the target state \emph{without going back to the source or going through the target in between}. Assuming that \emph{Sargassum} is transported in such a productive manner, we use TPT to connect a bloom that ``spans'' $B \subset \mathbb D$, observed at ``time'' $k = k^B$, with its origin at $k = 0$ ``within'' $\bar B = \mathbb D\setminus B$. (From this point onward, quotation marks are omitted, and we loosely equate domains with partitions, boxes with indices, and so on.)  This is achieved by analyzing transitions from $\omega$, considered as the source, to $B$, identified as the target.  By construction of the chain, \emph{such transitions are accomplished through $\bar B$}.  The origin of the bloom should then be delineated by the region(s) where most transitions emerge from $\bar B$, offering an alternative evaluation of the origin compared to Bayesian inference.

The main objects with which nonautonomous TPT \cite{Helfmann-etal-20} characterizes the above nonstationary transition paths are the \emph{time-dependent forward}, $q^+(k) = \smash{(q^+_i(k))_{i\in S}}$, and \emph{backward}, $q^-(k) = \smash{(q^-_i(k))_{i\in S}}$, \emph{committors}.  Specialized to the present setting, the $i$th forward committor gives the probability that, starting in $i \in S$ at $k \in \mathbb K$, the chain reaches first $B$ before $\omega$ within $\mathbb K$.  Namely, $q^+_i(k) := \Pr(T^+_B(k) < T^+_\omega(k) \mid X_k = i)$, where $T^+_A(k) := \inf_{n\ge k\in\mathbb K}\{t_n : X_n\in A\}$ is the \emph{first entrance} time to $A\subset S$ after or at $k\in\mathbb K$. This is evaluated by solving the iterative system of algebraic equations 
\begin{equation}
    \left\{
    \begin{aligned}
    &q^+_{i\in\bar B}(k) = \sum_{j\in S}P_{i\in\bar B,j}(k)q^+(k+1),\\ &q^+_{i\in\omega}(k) = 0,\\  &q^+_{i\in B}(k) = 1,
    \end{aligned}
    \right.
\end{equation}
for $k \in \mathbb K\setminus\!\{K-1\}$, with \emph{final} condition $q^+(K-1) = \chi_B(i)$. In turn, the $i$th backward committor gives the probability that, arriving in $i \in S$ at $k\in\mathbb K$, the chain leaves first $\omega$ after $B$ within $\mathbb K$.  Namely, $q^-_i(k) := \Pr(T^-_\omega(k) > T^-_B(k) \mid X_k = i)$, where $T^-_A(k) := \sup_{n\le k\in\mathbb K}\{t_n : X_n\in A\}$ is the \emph{last exit} time to $A\subset S$ before or at $k\in\mathbb K$. This is determined by iteratively solving 
\begin{equation}
    \left\{
    \begin{aligned}
    &q^-_{i\in\bar B}(k) = \sum_{j\in S}P^-_{i\in\bar B,j}(k)q^-_j(k-1),\\ &q^-_{i\in\omega}(k) = 1,\\  &q^-_{i\in B}(k) = 0,
    \end{aligned}
    \right.
\end{equation}
for $k \in \mathbb K\setminus\!\{0\}$, with \emph{initial} condition $q^-(0) = \chi_\omega(i)$.  Here, $P^-(k)$ is the transition matrix for the time-reversed chain $\{X^-_k\}$, which traverses the original chain backward in time, i.e., $X^-_k = X_{K-1-k}$.  Let $\lambda(0) \in \mathbb R^{1\times(N+1)}$ represent a probability vector at $k = 0$, i.e., such that $\sum_{i\in S} \lambda_i(0) = 1$.  At later times $k\in\mathbb K\setminus\!\{0\}$, $\lambda(k+1) = \lambda(k)P(k)$. Then \cite{Ribera-19}
\begin{equation}
    P^-_{ij}(k) := \Pr(X^-_{K-1+k+1} = j \mid X^-_{K-1+k} = i) = \frac{\lambda_j(k-1)}{\lambda_i(k)}P_{ji}(k-1)
    \label{eq:Pm}
\end{equation}
provided that $\lambda_i(k) > 0$\,$\forall i\in S$. This is quite different from autonomous TPT, wherein the reversed chain's transition matrix is determined by the stationary distribution of the direct chain, provided that this is ergodic and mixing, under the presupposition of equilibrium conditions.

The nonautonmous TPT \cite{Helfmann-etal-20} statistic of main interest to us is the \emph{time-dependent transition current} that, within our setting, provides at time $k \in \mathbb K$ the average flux of trajectories that traverse $i\in \mathbb S$ at time $k\in\mathbb K$ and subsequently pass through $j\in S$ at time $k + 1\in\mathbb K$, originating lastly from $\omega$ and subsequently advancing toward $B$, rather than $\omega$, viz., $f^{\omega B}_{ij}(k) := \Pr(X_k = i, X_{k+1} = j, T^-_\omega(k) > T^-_B(k), T^+_B(k+1) < T^+_\omega(k+1))$. The \emph{time-dependent effective transition current}, which represents the net transition current and minimizes detours, is captured by $f^+(k) = (f^+_{ij}(k))_{i,j\in S} \in \mathbb R^{(N+1)\times (N+1)}$ where 
\begin{equation}
    f^+_{ij}(k) := \max\big\{f^{\omega B}_{ij}(k) - f^{\omega B}_{ji}(k)\big\},
\end{equation} 
where $f^{\omega B}_{ij}(k) = q^-_i(k)\lambda_i(k) P_{ij}(k)q^+_j(k+1)$ for $k\in\mathbb K\setminus\!\{K-1\}$.

To visualize $f^+(k)$ on the partition $\mathcal D_N$ of the ocean surface domain $\mathcal D$, to each box $B_i\in\mathcal D_N$ one can attach at each time $t_k$ the vector on $\mathcal D_N$ given by 
\begin{equation}
    \mathbf f^+(\mathbf x_i,t_k) := \sum_{j\neq i\in S} f^+_{ij}(k)\mathbf e_{ij}(k),
    \label{eq:f}
\end{equation}
where $\mathbf e_{ij}(k)$ is the unit vector pointing from the center $\mathbf x_i$ of $B_i\in\mathcal D_N$ to the center of $B_j\in\mathcal D_N$, $j \neq i$.  This gives the magnitude and the direction of the effective transition current $f^+(k)$ out of each box $B_i\in\mathcal D_N$.  If the origin of a bloom in $\mathcal B\subset\mathcal D_N$ recorded at time $t_{k_B}$ is $\mathcal O \in\bar{\mathcal B}$, the expectation is that TPT will frame $\mathcal O$ as the region from which the (discrete) time-dependent vector field $\mathbf f^+(\mathbf x_i,t_k)$ is seen to emerge.   

\section{Bayesian and TPT Inferences of the 2011 Bloom's Origin}

\begin{figure}[t!]
    \centering
    \includegraphics[width=\linewidth]{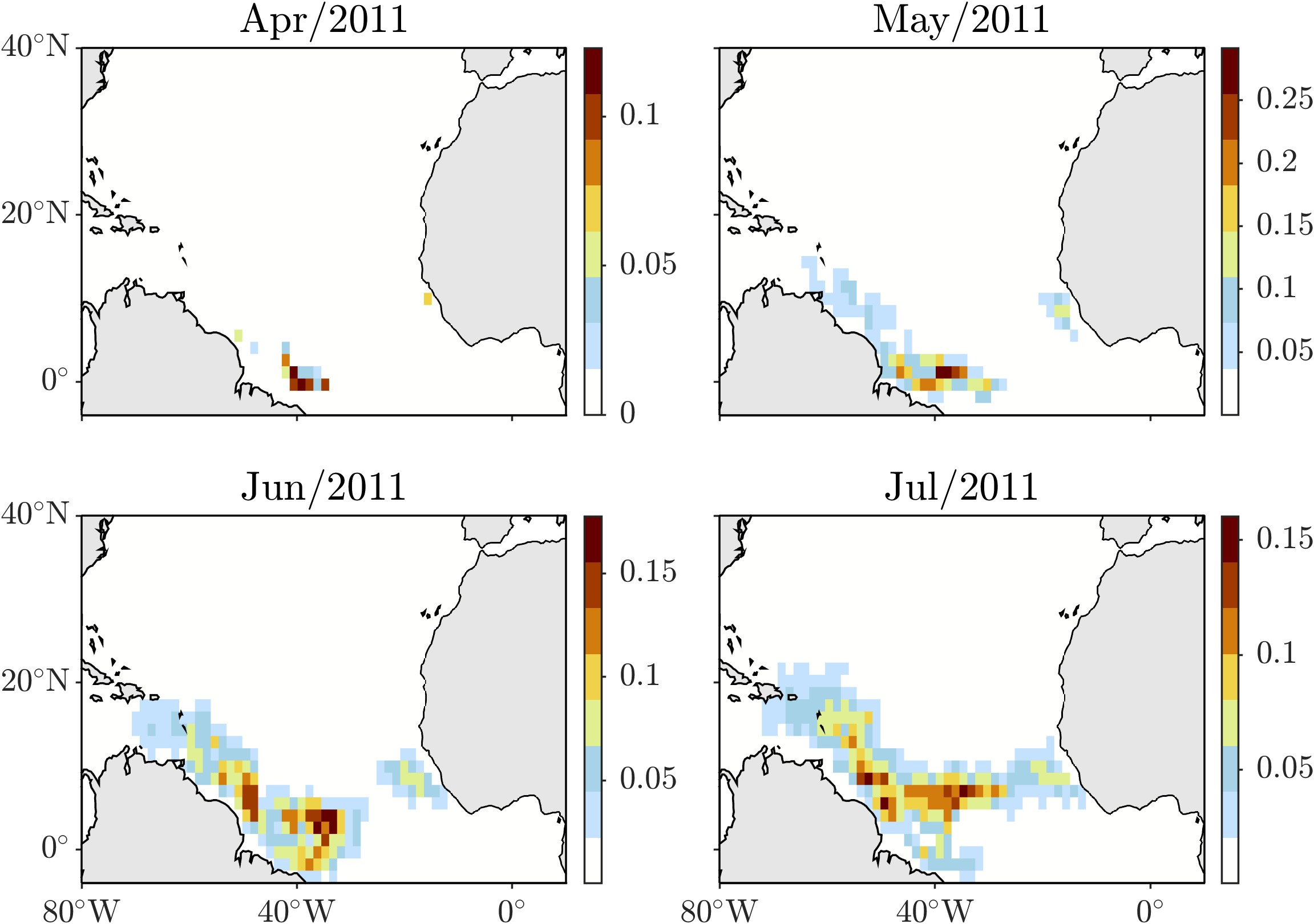}
    \caption{Forward evolution of the discrete representation of the probability density depicted in the top-left panel under left multiplication by the nonautonomous transition matrix \eqref{eq:Pext} constructed using \emph{Sargassum} clump trajectories obtained by integrating the eBOMB model equations \eqref{eq:eBOMB}, with ocean currents and winds as produced by reanalysis.  The initial density is randomly distributed on boxes intersecting with the first documented major \emph{Sargassum} bloom location in the tropical North Atlantic. A second-root transformation is applied to each distribution.}
    \label{fig:push}
\end{figure}

The focus of our analysis is on the first significant recorded bloom in the tropical North Atlantic, which occurred in April 2011. This bloom's location is indicated by the colored square patches in the top-left panel of Fig.\@~\ref{fig:push}. The patch locations were inferred from areas where the satellite-derived monthly cumulative density of \emph{Sargassum} was highest in April 2011, as shown in the animation included in the Supplementary Material of Wang et al.\@~\cite{Wang-etal-16}. The density of \emph{Sargassum} primarily peaks in a large western area of the tropical North Atlantic basin. However, there is also a small region of high \emph{Sargassum} density located off the southern coast of West Africa, specifically near Guinea. The other panels of Fig.\@~\ref{fig:push} show snapshots of the forward evolution of the probability density's discrete representation $d$, depicted in the top-left panel, representing a distribution over the boxes intersecting the observed bloom region. This forward evolution is achieved by left multiplying $d$ with the nonautonomous transition matrix $P(k)$, given in \eqref{eq:Pext}, constructed using \emph{Sargassum} clump trajectories that satisfy the eBOMB model equations \eqref{eq:eBOMB}. The ocean currents and winds are sourced from the ECMWF (European Centre for Medium-Range Weather Forecasts) ORAS5 (Ocean Reanalysis System 5) based on the OCEAN5 system \cite{Zuo-etal-19} and ERA5 (ECMWF Reanalysis v5) \cite{Hersbach-etal-20}, respectively. The ocean model in OCEAN5 features an eddy-permitting horizontal resolution of 0.25$^\circ$ and a near-surface vertical resolution of 1 meter, offering near real-time daily monitoring. ERA5 provides hourly global estimates on a 31 km grid, resolving the atmosphere with 137 levels from the surface up to 80 km. The eBOMB parameters were acquired through optimization, as reported in Table 2 of Bonner et al. \cite{Bonner-etal-24} This process resulted in a windage parameter value ($\alpha \approx 0.34$\%) consistent with laboratory measurements \cite{Olascoaga-etal-23}.  The integration is accomplished using the Julia code \texttt{Sargassum.jl}, which integrates eBOMB on the sphere and accounts for the effects of beaching and raft disaggregation.  The partition boxes have a side length of 1.5$^\circ$ and the transition increment $\Delta t = 1$\,week, which guarantees sufficient memory loss into the past for the Markovian assumption to hold approximately \cite{LaCasce-08}.  The transition matrix computation is based on the straightforward use of Matlab's \texttt{histcounts2.m}, which gave us full control on the grid size and location. This was feasible because our application does not require the assumption of ergodicity and the need of a dynamic grid adjustment as implemented in Julia's \texttt{UlamMethod.jl} and Python's \texttt{pygtm}. The evolution of the probability vector $d$ is in good qualitative agreement with the evolution of observed \emph{Sargassum} as documented in Wang et al.\@~\cite{Wang-etal-16}, supporting both the eBOMB model and the Markov chain reduction. A distinguishing feature is the development of a distribution resembling the GASB.  An important observation is that the omission of the box near the African coast in the initial distribution leads to an incomplete GASB-like distribution disconnected from the African coast, highlighting the importance of tracing the origin of the bloom there.

\begin{figure}[t!]
    \centering
    \includegraphics[width=\linewidth]{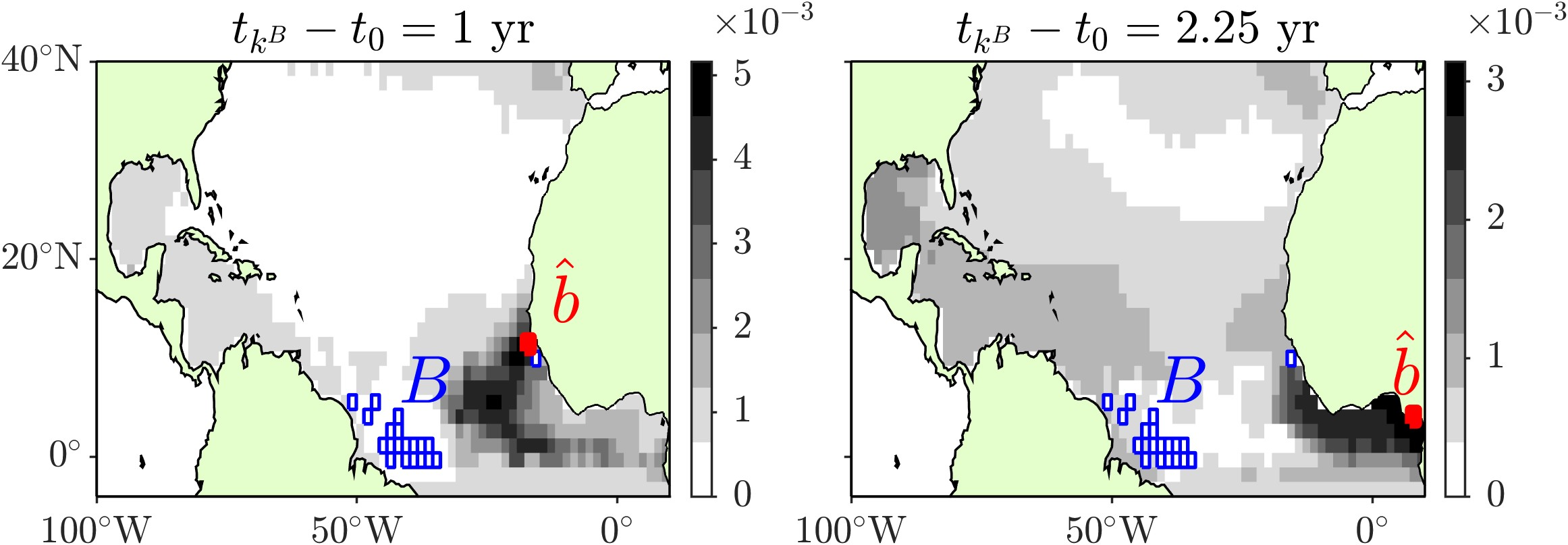}
    \caption{Posterior distribution of the location of the origin of the first documented major \emph{Sargassum} bloom intersecting with the ocean surface's negion partitioned by the set ($B$) of blue boxes, after the bloom has been observed in $t_{k^B} = \text{April}/2011$.  Results shown correspond to two initialization times $t_0$ of the time-inhomogeneous Markov chain, obtained via a reduction of the motion described by \emph{Sargassum} trajectories as simulated using eBOMB. Indicated in red is the maximum likelihood estimator of the location of the bloom’s origin, $\hat b$.}
    \label{fig:bayes}
\end{figure}

Figure \ref{fig:bayes} summarizes the results from the Bayesian inference of the bloom's origin. The set $B$ of box indices in this calculation are taken to be those of the square colored patches in the top-left panel of Fig.\@~\ref{fig:push}. Note that $B$ consists of a union of disconnected boxes, with most boxes situated closer to the South American coast, and one located off the coast of West Africa. Assuming that the knowledge of the bloom's origin before observing any data is nonexistent, we set $p(\bar{b}) = \text{const}$. The figure then shows the posterior distribution $p(\bar{b}\mid t_{k^B})$ for the location of the bloom's origin after it has been observed in $t_{k^B} = \text{April}/2011$. The posterior distribution is displayed for two initialization times $t_0$ of the Markov chain, corresponding to 1 year (left panel) and 2.25 years (right panel) prior to bloom observation. When $t_{k^B}-t_0 = 1$ year, $p(\bar{b}\mid t_{k^B})$ peaks in the western tropical North Atlantic. The maximum likelihood estimator of the origin of the bloom, marked by the red box labeled $\hat b$ in Fig.\@~\ref{fig:bayes}, lies near the western coast of West Africa, roughly corresponding to Guinea. This inference remains unaffected even if the target box directly south of this location is removed.  When $t_{k^B}-t_0 = 2.25$ years, corresponding to an initialization in February 2009, there is a significant increase in $p(\bar{b}\mid t_{k^B})$ within a region that includes the Gulf of Guinea, extending along the coast of West Africa from approximately the area corresponding to Guinea. The maximum likelihood estimator of the bloom's origin is situated within the Gulf of Guinea, near the coasts between Cameroon and Nigeria. For the 2009 initialization, $p(\bar{b}\mid t_{k^B})$ remains nonzero in regions such as the Intra-Americas Seas, the eastern side of the subtropical gyre, and a band south of the equator. However, these values are much smaller compared to those in the Gulf of Guinea.  Notably, in Miron et al.\@~\cite{Miron-etal-21-Chaos} and Beron-Vera et al.\@~\cite{Beron-etal-22-ADV}, the Gulf of Guinea was characterized, through an analysis using an autonomous Markov chain derived from satellite-tracked undrogued surface drifting buoys, as a region within the tropical North Atlantic that exhibits weak connectivity (high retention).

\begin{figure}[t!]
    \centering
    \includegraphics[width=\linewidth]{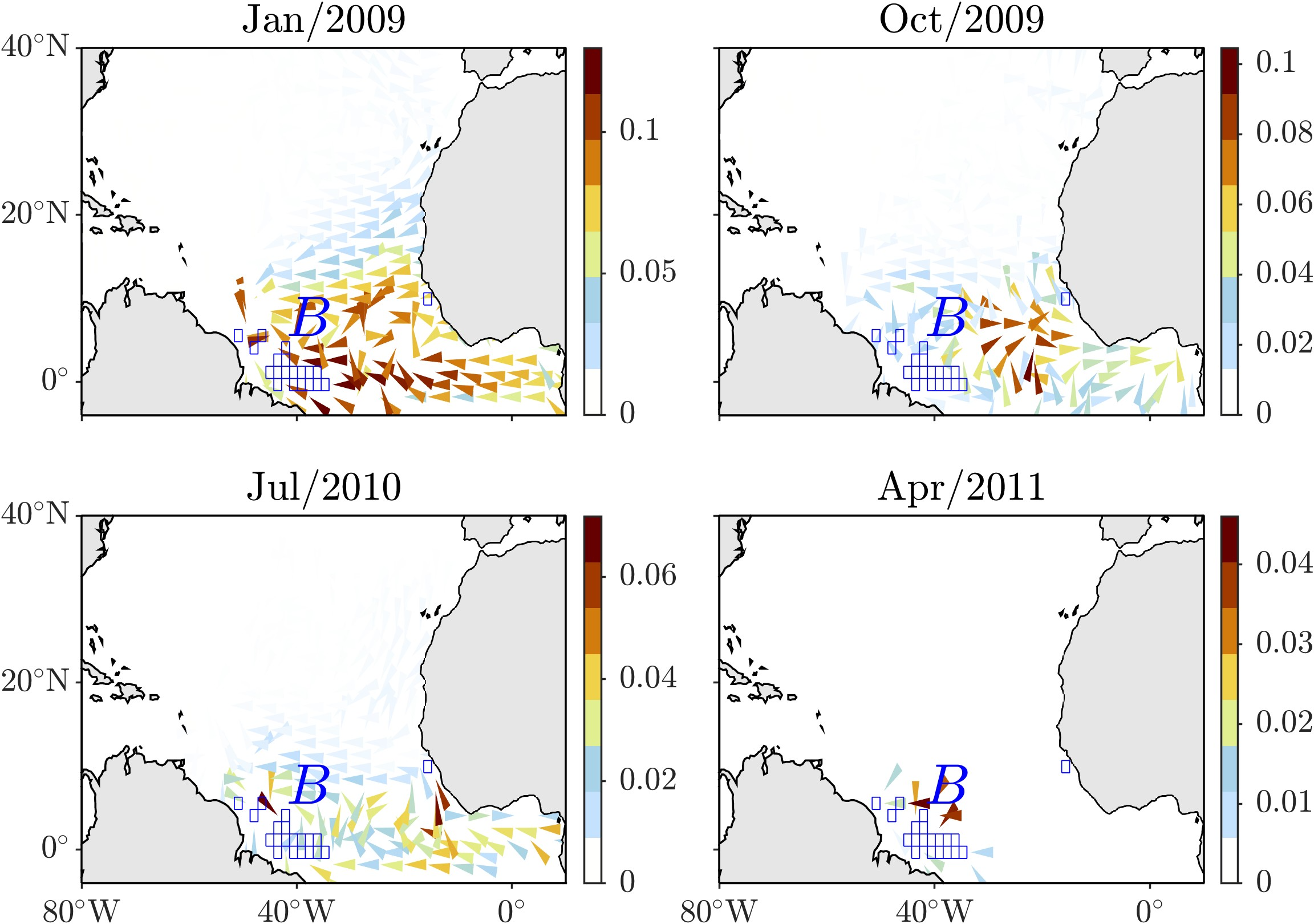}
    \caption{Snapshots of time-dependent effective transition currents into the region spanned by the \emph{Sargassum} bloom observed in April/2011, for an initialization 2.25 years before this observation of the time-inhomogenous Markov chain constructed by discretizing \emph{Sargassum} trajectories as produced by eBOMB. A fourth-root transformation is applied to the current magnitude.}
    \label{fig:tpt}
\end{figure}

The results from the TPT inference are shown in Fig.\@~\ref{fig:tpt}. This figure presents several snapshots of time-dependent effective transition currents entering the set $B$ of boxes that intersect with the bloom region. These currents result when the Markov chain is initialized in February 2009, 2.25 years before the April 2011 bloom. This allows for a comparison with the right panel of Fig.\@~\ref{fig:bayes}. The emergence of transition currents is observed within a vast area of the tropical North Atlantic, covering the target set $B$, and stretching from the equator to approximately 10$^\circ$N. This area extends from the coast of South America to the coast of West Africa, including the Gulf of Guinea. By April 2011, all currents had progressed to $B$. Between February 2009 and April 2011, they experienced a bottleneck toward the eastern portion of the tropical North Atlantic. The currents do not suggest a direct transition path to $B$, although they represent paths that aim to minimize detours. As with Bayesian inference, the configuration of the transition currents remains unchanged even if the box of the target set located near the African coastline is excluded from the analysis.  These results are consistent with direct trajectory integrations \cite{Franks-etal-16, Putman-etal-18, Alleyne-etal-23, Gobert-etal-25} suggesting southern zonal pathways similar to those inferred from the present TPT analysis. Moreover, Fidai et al.\@~\cite{Fidai-etal-25} supports the idea that beaching on the coasts of West Africa is on the rise. This indirectly substantiates the idea of recirculation there, consistent with prior autonomous Markov-chain analyses \cite{Miron-etal-21-Chaos, Beron-etal-22-ADV}, and subsequent westward transport, as reported here.

In comparing the Bayesian and TPT inferences, it becomes evident that the former provides a more precise determination of the bloom's origin near the coast of West Africa than the latter. Nevertheless, it is important to acknowledge that the conclusions drawn from both the Bayesian source inversion and the TPT assessment of the bloom's genesis are exclusively based on physical constraints. Consequently, it is necessary for these findings to be substantiated by biological constraints. This verification will be addressed in the subsequent section following the introduction of a significant observation presented through an appropriate adaptation of TPT.

\begin{figure*}[t!]
    \centering
    \includegraphics[width=\textwidth]{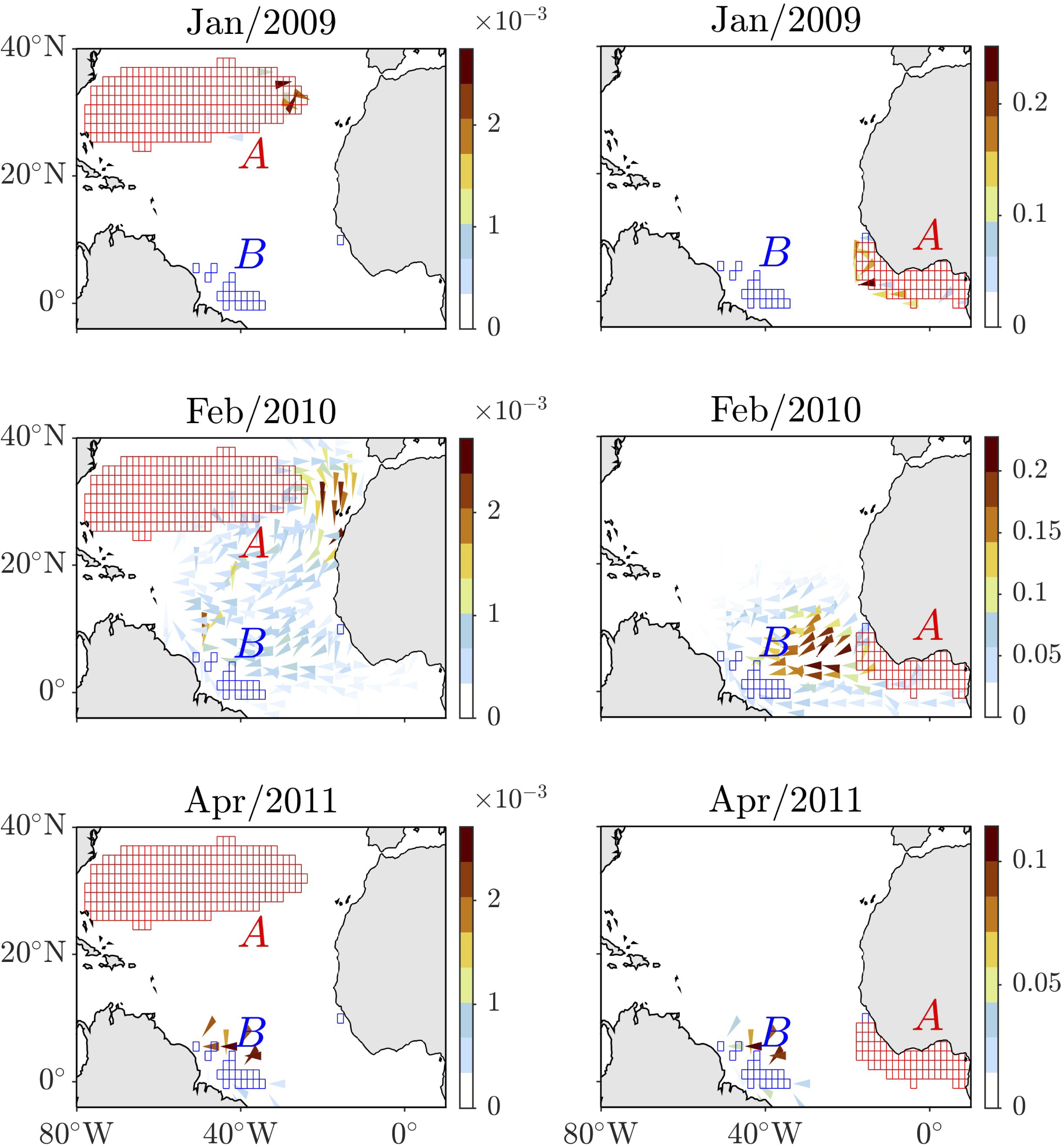}
    \caption{Similar to Fig.\@~\ref{fig:tpt}, but with the source set ($A$) spanning the Saragasso Sea (left panels) and a region including the the Gulf of Guinea (right panels), as informed by Bayesian inference of the bloom's origin.  The latter is informed by Bayesian inference, highlighting the top 50th percentile of the posterior distribution for the bloom's origin, as depicted in Fig.\@~\ref{fig:bayes}, right panel.}
    \label{fig:tpt-SS}
\end{figure*}

Previous studies \cite{Johns-etal-20, Jouanno-etal-25} have identified that the origin of \emph{Sargassum} blooms in the tropical North Atlantic is within the Sargasso Sea rather than nearshore West Africa.  While the analysis above does not support this possibility, we further test its validity using TPT with adjusted source set placement. Specifically, this involved changing from $\omega$---the virtual state used to close the Markov chain---to a set of boxes ($A$) strategically selected to cover a region across the Sargasso Sea (subtropical gyre) and a region of the tropical North Atlantic off the coast of West Africa, as suggested by Bayesian inference of the bloom origin. The latter corresponds to the boxes representing the top 50th percentile of the posterior distribution of the bloom's origin location after it was observed in April 2011, as shown in the right-hand side panel of Fig.~\ref{fig:bayes}. This corresponds to the initialization of the Markov chain in February 2009. The TPT formulae remain unchanged, except for replacing $\omega$ with $A$. The results are shown in Fig.~\ref{fig:tpt-SS}. As in Fig.~\ref{fig:tpt}, the initialization is set 2.25 years before the observed bloom time. Although a transition current does originate from the Sargasso Sea, its magnitude is significantly smaller---up to eight orders of magnitude less---than the considerably more robust current emanating from the coastal waters of West Africa (note that a fourth-root transformation has been applied to the current magnitude). Such a minor transition current into the bloom site appears comparatively insufficient to effectively transport \emph{Sargassum} to stimulate a bloom, challenging the assertion made in Johns et al.\@~\cite{Johns-etal-20} and Jouanno et al.\@~\cite{Jouanno-etal-25}, given the considerably stronger currents from offshore West Africa. 

\section{Environmental Conditions Favoring Bloom Development}

Despite the anecdotal report \cite{Addicod-deGraft-16} of \emph{Sargassum} rafts landing on the shores of Ghana, and the consistent constraints imposed by Bayesian inversion and TPT analysis on the bloom's origin in the nearshore waters of West Africa, a persistent question concerns the mechanism of bloom initiation. Considering the low probability of \emph{Sargassum} being transported into this area, it is hypothesized that a pre-existing baseline concentrations of \emph{Sargassum} within the region evolved into a bloom, facilitated by favorable environmental conditions produced by a strong ``coastal Ni\~na'' event. 

\paragraph{Biogeochemistry Hindcast.}

To explore this hypothesis, we analyze daily outputs from comprehensive global biogeochemistry hindcast GLOBAL\_MULTIYEAR\_BGC\_001\_029, based on the FREEBIORYS2V4 simulation utilizing PISCES-v2 (Pelagic Interactions Scheme for Carbon and Ecosystem Studies, version 2) \cite{Aumont-etal-15}. The simulation is forced by daily-averaged fields from the ocean, sea ice, and atmosphere. Oceanic and sea ice forcings are provided by FREEGLORYS2V4, which employs NEMOv3.1 (Nucleus for European Modelling of the Ocean, version 3.1) \cite{Madec-16}, while atmospheric forcings are derived from the ECMWF ERA-Interim reanalysis \cite{Dee-etal-11}. The biogeochemical model includes monthly climatologies of atmospheric dust deposition for nitrate, phosphate, silicate, and iron, as well as riverine inputs aligned with runoff locations in the physical model. The simulation is initialized in December 1991 using nutrient fields (nitrate, phosphate, oxygen, and silicate) from the World Ocean Atlas 2013 \cite{Garcia-etal-14a, Garcia-etal-14b} and carbon data from GLODAPv2 (Global Ocean Data Analysis Project, version 2) \cite{Lauvset-etal-22}. After a spin-up period, the simulation spans from 1993 to two months before the present, with a horizontal resolution of 0.25$^\circ$. However, a steady-state equilibrium is only reached around the year 2000.

\begin{figure*}[t!]
    \centering
    \includegraphics[width=\textwidth]{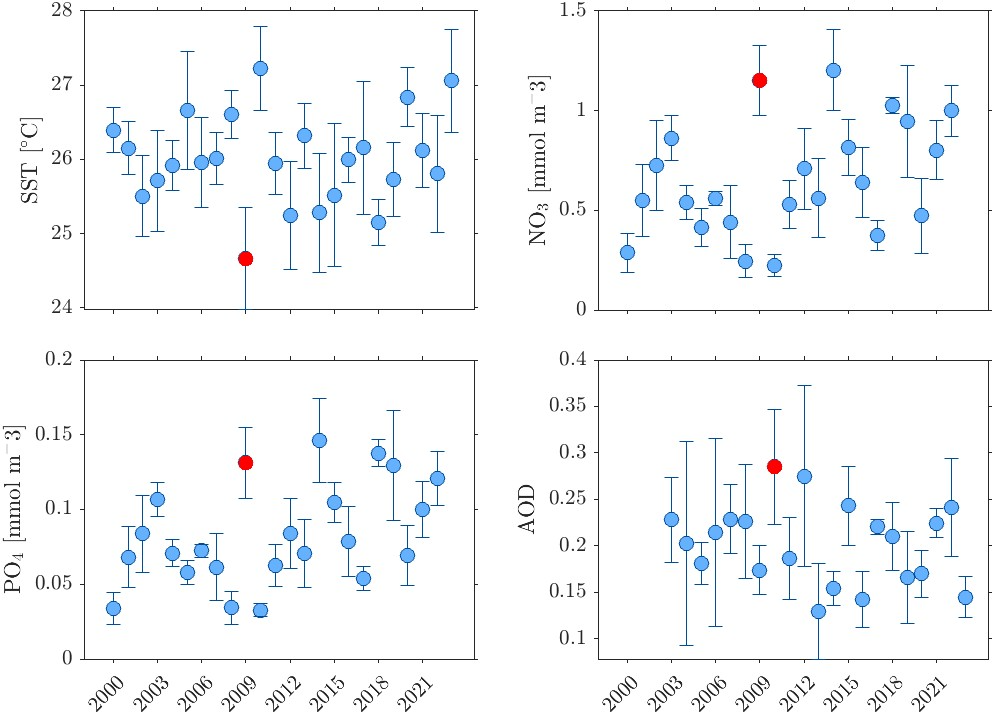}
    \caption{Sea-surface temperature (top-left), nitrate (top-right), and phosphate (bottom-left), as produced by a biogeochemical hindcast, and dust extinction aerosol optical depth (bottom-right), as produced by a reanalysis system, all averaged within [21$^\circ$W,17$^\circ$W] $\times$ [9$^\circ$N,14$^\circ$N] in spring.}
    \label{fig:env}
\end{figure*}

Within this framework, we focus on sea-surface temperature (SST), phosphate (PO$_4$), and nitrate (NO$_3$) in the period 2000---2023, averaged over the region off Senegal bounded by [21$^\circ$W,17$^\circ$W] $\times$ [9$^\circ$N,14$^\circ$N]. This region is significant as it includes the maximum likelihood estimate of the bloom's origin, initialized a year before the bloom observation (Fig.\@~\ref{fig:bayes}, left panel). The selection of this region is further supported by a recent independent study by Oettli et al.\@~\cite{Oettli-etal-16}, which will be cited later to substantiate the biogeochemical simulation. Resultant time series data restricted to spring are depicted in Fig.\@~\ref{fig:env}, in the top-left, top-right, and bottom-left panels, respectively. Each box plot illustrates the mean values with error bars corresponding to one standard deviation around them. The box representing the year 2009 is highlighted in each panel, showing substantially higher values of PO$_4$ and NO$_3$ compared to preceding years, while SST reached a minimum in 2009. Elevated concentrations of nitrate and phosphate are favorable for bloom development \cite{Lapointe-etal-21}. The effect of temperature is not well understood. Pelagic \emph{Sargassum} species can survive in temperatures between 18 and 30$^\circ$C \cite{Hanisak-Samuel-87}. However, there is no consensus on the optimal conditions for their growth \cite{Hanisak-Samuel-87, Corbin-etal-23, Magana-etal-23b}.

Oettli et al.\@~\cite{Oettli-etal-16} investigated the interannual variability of coastal SST anomalies off Senegal, similar to patterns developed during ``coastal Ni\~no/Ni\~na'' in the Northeastern Pacific and Southeastern Indian Oceans. Warm (cold) events were dubbed Dakar Ni\~no (Ni\~na) by Oettli et al.\@~\cite{Oettli-etal-16}. A Dakar Ni\~no (Ni\~na) involves anomalous warming (cooling) of the thin (thick) mixed-layer due to heat absorption (release). For Dakar Ni\~na, entrainment through the mixed-layer's bottom is significant. Oettli et al.\@~\cite{Oettli-etal-16} demonstrated, using reanalyzed SST data, that the standardized coastal SST index for the area, which was utilized to construct Fig.\@~\ref{fig:env}, shows a very pronounced minimum in 2009. This classifies the year as a strong Dakar Ni\~na, explaining the anomalous SST, nitrate, and phosphate levels observed in the biogeochemical simulation.

\paragraph{Dust and Land Runoff.}

An additional environmental condition examined to support the hypothesis is the amount of dust and other particulate matter in the atmosphere, with the Sahara Desert in northern Africa being a significant source. This is commonly measured by the dust extinction aerosol optical depth (AOD) at a 550 nm reference wavelength, which is a unitless measure quantifying the amount of light blocked by dust particles in the atmosphere. We analyzed this parameter as produced monthly by EAC4 (ECMWF Atmospheric Composition Reanalysis 4), a global reanalysis of atmospheric composition \cite{Inness-etal-19}. Changes in the observing system, despite assimilation systems being capable of resolving data gaps, initially resulted in sparser networks and less accurate estimates, limiting EAC4 availability to post-2003. The bottom-right panel of Fig.\@~\ref{fig:env} illustrates AOD in a manner consistent with the other panels; however, it uses monthly data. AOD is observed to reach its maximum in 2010. If elevated AOD is taken as a proxy for increased atmospheric nutrient deposition to the ocean, this could suggest a mechanism supporting the persistence of \emph{Sargassum} blooms beyond 2009. Analysis of dust data from 1973 to 2017 in Barbados reveals a significant decline in dust concentrations from 1982 to 2008, a trend observed consistently across all four seasons \cite{Zuidema-etal-19}. However, this decline does not persist into the present decade. In particular, spring-–summer dust mass concentrations have shown increased interannual variability. Since 2013, most summers have experienced above-average dust loads, indicating a departure from the earlier long-term decline. Additionally, Gaston et al.\@~\cite{Gascon-etal-24}, using dust data from 1990 to 2011 from Barbados as well, reported that nitrate concentrations remained relatively stable throughout the period, with the exception of notable elevations during the spring of 2010 and the summer and fall of 2008. An important observation is that the biogeochemistry model hindcast discussed above was driven only by climatological dust deposition, and therefore does not account for interannual variability in atmospheric inputs such as the 2010 AOD peak.

Another significant source of nutrients that deserves to be mentioned is the runoff of land from rivers. The biogeochemical model only accounts for the river input through climatological averages. However, extreme flooding events, such as those reported in the Senegal and Niger River basins in 2009 and 2010 \cite{Wilcox-etal-18, Ndiaye-etal-23}, can lead to significant episodic increases in discharge into the ocean, resulting in nutrient influx.

\paragraph{Baseline \emph{Sargassum} Concentrations.}

The suggestion that baseline concentrations of \emph{Sargassum} existed in the tropical North Atlantic, including regions off West Africa, has yet to be validated. This assumption forms a cornerstone of the Markov model developed in this study and supports our broader hypothesis. Historical evidence indicates that Sargassum was present in the tropical North Atlantic before 2011, albeit in smaller quantities, primarily as \emph{S.\@~natans} VIII. Parr \cite{Parr-39} documented the occurrence of \emph{S.\@~natans} VIII in the southeastern Caribbean, seeing this as indicative of its westward spread from the tropical North Atlantic through the North Equatorial Current. Along the Brazilian coastline, \emph{S.\@~natans} was initially documented in the floristic inventory compiled by Taylor \cite{Taylor-31}. In recent years, de~Szechy et al.\@~\cite{deSzechy-etal-12} noted that Oliveira Filho, using unpublished data in an assessment of Brazilian macroalgal species, identified \emph{S.\@~natans} as a species of uncertain status but acknowledged its sporadic presence in northern and northeastern Brazil. The presence of \emph{S.\@~natans} was reported near the West African coast, including Gabon \cite{John-etal-04}, and the Cape Verde Islands \cite{Price-etal-78, John-etal-04, Prudhomme-etal-05}. For a comprehensive list, see Price et al.\@ \cite{Price-etal-78}. Furthermore, satellite-derived time series of \emph{Sargassum} abundance beginning around 2000 (cf., e.g., Fig.\@~4 in Gower et al.\@~\cite{Gower-etal-13}) confirm its persistent, albeit low, presence in the tropical North Atlantic during that period.  In sum, there is enough evidence supporting the hypothesis that \emph{Sargassum} indeed existed in baseline concentrations in the tropical North Atlantic, including regions off West Africa.

\paragraph{Summary and Discussion.}

This study simulates the movement of \emph{Sargassum} rafts by modeling them as systems of finite-size floating particles (clumps) subjected to Maxey--Riley dynamics and nonlinear elastic interactions. Surface ocean currents and wind data from reanalysis systems were utilized to define the flow transporting these clumps and to compute trajectories across the tropical and subtropical North Atlantic surface ocean. The motion described by these trajectories was then reduced through the application of Ulam's method, resulting in a time-inhomogeneous Markov chain. The openness of the flow domain introduces an imbalance in the probability mass along the chain, which was addressed by incorporating a virtual state that absorbs this imbalance and redistributes it uniformly back into the chain, thus simulating the presence of a background concentration of \emph{Sargassum}. Subsequently, Bayesian inversion, combined with nonautonomous transition path theory---a probabilistic approach enabling the rigorous characterization of nonequilibrium productive communication channels within the flow---was employed to determine the origin of the first significant bloom in the tropical North Atlantic, recorded in April 2011 on the basin's western side. Both methodologies consistently identified coastal West African waters as the source of the bloom. Our results are consistent with the \emph{Sargassum} strandings anecdotally reported along the shores of Ghana in 2009, aligning with and the observation of anomalously high environmental parameters associated by a strong Dakar Ni\~na event. These conditions are conducive to bloom stimulation in the presence of \emph{Sargassum} species in the baseline concentrations characteristic of the 2011 bloom. 

The results of our research provide a more coherent perspective compared to earlier ``leeway'' modeling studies \cite{Johns-etal-20, Jouanno-etal-25}, which identified the origin of \emph{Sargassum} blooms within the subtropical North Atlantic, specifically in the Sargasso Sea. A crucial element that supports our findings is that, in the onset of the tropical blooms, the dominant species (morphtype) was \emph{S.\@~natans} VIII. This can be inferred from a number of references \cite{Schell-etal-15, Govindarajan-etal-19, Oxenford-etal-21, Alleyne-etal-23b, Corbin-Oxenford-23}, for the lacking of direct morphtype analyses of 2011 \emph{Sargassum samples}. By contrast, \emph{S.\@~natans} I and \emph{S.\@~fluitans} III were prevalent in the Sargasso Sea \cite{Parr-39, Schell-etal-15, Govindarajan-etal-19}.  Moreover, the \emph{Sargassum} strandings in the Gulf of Guinea, as noted by Addicod and de~Graft \cite{Addicod-deGraft-16}, occurred before the extreme NAO event in 2009--2010, which has been linked to the connectivity between the Sargasso Sea and the tropical North Atlantic \cite{Johns-etal-20, Jouanno-etal-25}. Although a small amount of \emph{Sargassum} could have been transported from the Sargasso Sea to the tropical North Atlantic, differences in bloom characteristics between the tropics and subtropics---such as dominant morphtypes and abundance---still need further consideration. 

Several authors \cite{Howard-Menzies-69, Carpenter-Cox-74, Mann-etal-80} have previously indicated low productivity of \emph{Sargassum} in the Sargasso Sea. However, this view was revised following direct growth measurements of pelagic \emph{Sargassum} \cite{Lapointe-86, Hanisak-Samuel-87}, which reported that \emph{Sargassum} can double its biomass in just a few days or weeks under optimal conditions. Physiological studies \cite{Lapointe-95} and seawater nutrient analyses have shown that increased nitrogen and phosphorus availability significantly elevates the productivity of the macroalga in neritic compared to oceanic waters of the western subtropical North Atlantic. By comparing samples collected before 2010 with more recent ones, Lapointe et al.\@~\cite{Lapointe-etal-21} concluded that increased nitrogen availability is likely fueling large-scale \emph{Sargassum} blooms. They also noted that, given the high nitrogen\,:\,phosphorus (N\,:\,P) ratio of \emph{Sargassum} currently found in the North Atlantic basin, an excess of soluble reactive phosphorus in upwelled waters could stimulate further growth in the eastern tropical North Atlantic. More recently, McGillicuddy et al.\@ \cite{McGillicuddy-etal-23} noted that the nutritional status of \emph{Sargassum} in the tropics (the GASB) is different, with higher nitrogen and phosphorus contents than populations in the subtropical habitat of the Sargasso Sea. Changeux et al.\@~\cite{Changeux-etal-23} from in situ short-term experiments on Martinique Island concluded that differences in growth and tissue composition suggest that \emph{S.\@~fluitans} III was favored by conditions in the western tropical coastal North Atlantic. This raises a key question: why was \emph{S.\@~natans} VIII---rather than \emph{S.\@~natans I and/or} \emph{S.\@~fluitans} III---the dominant species at the onset of the GASB? A likely explanation is that the background concentration of \emph{S.\@~natans} VIII in the region was higher than the amount of \emph{S.\@~natans I} and/or \emph{S.\@~fluitans} III that could have been transported from the Sargasso Sea.

In conclusion, our results suggest that the 2011 bloom was driven by distinct local environmental conditions---including elevated nutrient availability in the tropical North Atlantic due to upwelling  off  West Africa and Saharan dust deposition there---as well as by regional ocean circulation dynamics.

\section{Acknowledgments}

We thank Ligia Collado (Florida International University) and Susana Enr\'iquez (Universidad Nacional Aut\'onoma de M\'exico) for their helpful discussions on the morphology and physiology of \emph{Sargassum}, Philip Tuchen (Cooperative Institute For Marine And Atmospheric Studies) for calling our attention to Dakar Ni\~na, and Philippe Cecchi (Universit\'e de Montpellier, France), Hassan Moustahfid (National Oceanic and Atmospheric Administration), Kafayat Fakoya (Universit\'e de Montpellier), and Ricardo Tabraue (Universidad de Las Palmas de Gran Canaria) for providing bibliographical references.

\section{Supplementary Material}

Supplementary material is available at PNAS Nexus online.

\section{Funding}

Our work was funded by the National Science Foundation (NSF) under grant OCE2148499.

\section{Author contributions statement}

F.J.\ Beron-Vera: Conceptualization; Formal analysis---Bayesian \& TPT inferences; Interpretation of the results; Writing---main text; review \& editing; Funding acquisition. M.J.\ Olascoaga: Conceptualization; Formal analysis---eBOMB trajectory integrations \& environmental data analysis; Interpretation of the results; Writing---review \& editing; Funding acquisition.  P.\ Miron: Interpretation of the results; Writing---review \& editing.  G.\ Bonner: Interpretation of the results; Code development; Writing---review \& editing. 

\section{Data availability}

The Julia package \texttt{Sargassum.jl} is available to the community from \href{https://github.com/70Gage70/Sargassum.jl}{https://\allowbreak github.com/\allowbreak 70Gage70/\allowbreak Sargassum.jl}.  The ECMWF's ORAS5 reanalysis data are distributed via \href{https://www.ecmwf.int/en/forecasts/dataset/ocean-reanalysis-system-5}{https://\allowbreak www.ecmwf.int/\allowbreak en/\allowbreak forecasts/\allowbreak dataset/\allowbreak ocean-reanalysis-system-5/}. The wind velocity data used originate from the ECMWF's ERA5 reanalysis, which can be accessed via \href{https://www.ecmwf.int/en/forecasts/dataset/ecmwf-reanalysis-v5}{https://www.ecmwf.int/en/forecasts/dataset/ecmwf-reanalysis-v5}.  The GLOBAL\_MULTIYEAR\_BGC\_001\_029 biogeochimical simulation is available from \href{https://doi.org/10.48670/moi-00019}{https://\allowbreak doi.org/\allowbreak 10.48670/\allowbreak moi-00019}. Finally, EAC4 data can be retrieved from \href{https://ads.atmosphere.copernicus.eu/datasets/cams-global-reanalysis-eac4?tab=overview}{https://\allowbreak ads.\allowbreak atmosphere.\allowbreak copernicus.\allowbreak eu/\allowbreak datasets/\allowbreak cams-global-reanalysis-eac4?tab=\allowbreak overview}.

\bibliographystyle{abbrvnat}

\end{document}